\documentclass[iop,apj,numberedappendix,twocolappendix]{emulateapj_cmcnally}
\usepackage{amsmath}
\usepackage{bm}
\usepackage{times}
\newcommand{\ez}{{\bm{\hat{e}}_z}}

\begin{document}

\title{Photophoresis in a Dilute, Optically Thick Medium and Dust Motion in Protoplanetary Disks}
\shorttitle{Photophoresis in a Dilute, Optically Thick Medium }

\author{Colin P.~McNally\altaffilmark{1}}

\author{Alexander Hubbard\altaffilmark{2}}

\altaffiltext{1}{Niels Bohr International Academy, Niels Bohr Institute, University of Copenhagen, Blegdamsvej 17, 2100 Copenhagen \O, Denmark}
\email{{\tt cmcnally@nbi.dk}}
\altaffiltext{2}{American Museum of Natural History, New York, NY, USA}
\email{{\tt ahubbard@amnh.org}}

\begin{abstract} 
We derive expressions for the photophoretic force on opaque spherical particles in a dilute gas
in the optically thick regime where the radiation field is in local thermal equilibrium.
Under those conditions, the radiation field has a simple form, leading to 
well defined analytical approximations for the photophoretic force that also consider both the internal thermal conduction within
 the particle, and the effects of heat conduction and radiation to the surrounding gas. 
We derive these results for homogeneous spherical particles; and for the double layered spheres appropriate 
for modeling solid grains with porous aggregate mantles.
Then, as a specific astrophysical application of these general physical results, we explore the parameter 
space relevant to the photophoresis driven drift of dust in protoplanetary disks. We show that highly porous silicate grains have sufficiently
low thermal conductivities that photophoretic effects, such as significant relative velocities between particles with differing porosity 
or levitation above the midplane, are expected to occur.
\end{abstract}


\section{Introduction}

Photophoresis \citep{1918AnP...361...81E} is a set of phenomena where
an incident flux of light drives particle motion though a gas by interaction with the gas molecules.
In so-called `$\Delta T_s$-photophoresis', or thermo-photophoresis, illumination results in
 a thermal gradient across the grain.
As a result, gas molecules recoil from the hot side of the dust grain more violently than
they bounce off the cold side, generating a net force.
Given sufficiently uniform surface properties and thermal conductivity within the particle, the
side of the particle 
that sees a greater incident light flux
will be hotter than the opposite side, even in the face of
internal thermal conductivity within the particle and thermal conduction to the surrounding gas.
In this scenario
positive photophoresis results, 
pushing the dust grain away
from the source of illumination. 

Most of the photophoresis literature in astrophysics covers the case of direct illumination, 
which only formally applies in an optically thin medium.
 This is also the case for literature discussing
 photophoresis in the context of protoplanetary disk, which has to date
 primarily focussed on the transport of dust grains in optically thin regions of the disk illuminated by
 the proto-star. However, many sections of protoplanetary disks and other astrophysical 
 environments are optically thick, so the basic formulation must be extended to cover those environments.
 
 Moreover, recent work has shown that turbulent energy dissipation in the accretion disk
 context is highly localized, resulting in strong spatial temperature fluctuations insulated by
 high optical depths \citep{2014ApJ...791...62M}. Such hot spots are associated with radiative fluxes which drive photophoresis forces:
 in effect, a dust grain sees, not its immediate environment, but rather a bubble approximately an optical
 depth in radius whose surface temperature can vary meaningfully.
 The optically thick regime of photophoresis is the 
 appropriate theory for deriving the forces and drift speeds of dust in such an environment.

 Attention was drawn to the to the possible role of photophoresis in protoplanetary disks by \citet{2005ApJ...630.1088K}, 
 focusing primarily on the late, optically thin, stages of protoplanetary disk evolution.
Photophoretic drift of particles in a protoplanetary disk has been studied in the context of disks which are sufficiently
 transparent for the direct illumination from the star to be dominant at the midplane \citep{2007A&A...466L...9M,2011A&A...531A.106M}.
\citet{2009ASPC..414..509W} examined photophoresis driven dust levitation in a protoplanetary disk, but 
applied the optically thin approach with a radiation source located at the disk midplane,
and in \citet{2009M&PS...44..689W} proposed that photophoresis can levitate
CAIs above the disk in the optically thin region during an FU~Ori outburst.
In a work immediately preceding this one, \citet{paper1}, the scope for photophoresis was expanded
 to temperature fluctuations within the body of a protoplanetary disk, albeit in a formally one-dimensional treatment. 
In the atmospheres of accreting giant planets, photophoresis, even in the optically thick regime, 
can levitate and push out dust grains, limiting the opacity and accelerating the formation of these planets \citep{2013A&A...555A..98T}.
 
  The  dust grain's thermal conductivity is a central parameter in determining the photophoretic force. 
 High conductivity reduces the internal thermal gradients and hence the effectiveness of photophoresis; and existing
 studies of photophoresis in protoplanetary disks considered specifically the difference between silicate and metal
 dust grain conductivities \citep{2013ApJ...769...78W}.
 Solid particles which acquire a porous mantle of uncompacted dust with low conductivity will experience enhanced photophoresis
 \citep{2012A&A...545A..36L}.
 
  In this paper, we extend the optically thin direct illumination framework of photophoresis to the optically thick limit,
 including the effect of thermal conduction with the ambient gas as well as the impact of concentric shells
 with different thermal conductivities. 
Gratifyingly, all these disparate factors can be wrapped into a simple and powerful analytical closed form for the
photophoretic forces and drift velocities.
As an application, we consider dust motions in protoplanetary disks
and show that the vastly differing thermal conductivities of highly porous and highly compacted dust grains 
cause a meaningful distinction between the effectiveness of photophoresis for those grains.
 In inner regions of protoplanetary disks, we find that photophoresis can dominate over radial drift as a driver of collisions.
 Finally, we suggest that photophoresis driven by the release of accretion energy deep in the disk can levitate
  particles to several scale heights in a protoplanetary disk.

\section{Theory of Photophorersis in a Dilute, Optically Thick Medium}
\label{sec_theory}

We concern ourselves with so-called `$\Delta T$-photophoresis', or thermo-photophoresis, arising from variation in the particle surface 
temperature $T_s$; as opposed
 to a photophoresis force resulting from any variation over the surface of the thermal accommodation coefficient $\alpha$
 (which measures the rate with which gas molecules thermally equalize with the surface as they impinge upon it).
 We also restrict our attention here to opaque particles 
 { with a size greater than the wavelength corresponding to the temperature of the radiation field}, 
 so the photophoresis is positive (in the direction of the net light flux).
 Furthermore the theory that we derive shall be applicable in the limit that the particle surface temperature is close to the surrounding gas temperature.
 { Finally, for simplicity, we assume perfect blackbody absorptivities and emissivities. Non-perfect blackbodies will have equally weaker
 radiation heating and cooling terms, increasing the relative effectiveness of the internal thermal conductivity and heat conduction to the gas.}
 
In the free molecular flow regime, which we term `dilute' and which is
characterised by a Knudsen number $>1$, the photophoresis force is generated by the momentum of particles impinging on the surface and re-emerging \citep{rohatscheck1985,rohatscheck1995}:
\begin{align}
{ F}_p &= -\frac{1}{2} \oint_{S} p \left( 1+ \sqrt{\frac{\tilde{T}}{T_g}} \right) d{S}, \label{eq_photforce_general}\\
\tilde{T} &\equiv T_g + \alpha (T_s-T_g),
\end{align}
where $p$ is the background pressure, $T_g$ is the background gas temperature, $\alpha$ is the thermal accommodation coefficient of the particle surface 
at temperature $T_s$, and the integral is taken over the surface of a particle.
Of the above quantities, $T_s, T_g, p$, and $\alpha$, we assume that only $T_s$ (and hence $\tilde{T}$) varies across the particle's surface.

As in \citet{rohatscheck1985} we note that the position-independent part of the integral does not contribute, and linearize the square root in the limit 
\begin{align}
\zeta \equiv \frac{\alpha}{2}(T_s/T_g-1) \ll 1. \label{zeta_definition}
\end{align}
Expanding, to linear order we have
\begin{align}
\sqrt{1+ 2\zeta } = 1+\zeta + \mathrm{O}\left(\zeta^2\right).
\label{root_zeta_expansion}
\end{align}
Because $\alpha$ is a constant of order unity, in practice this approximation is  that $T_s/T_g-1 \ll1$, 
i.e.~that the particle surface temperature is close to the surrounding gas temperature.
To help sustain that approximation, we restrict ourselves to systems with sufficiently slowly spatially
varying gas temperature
that the temperature field can be treated as linear out to several optical depths. 

We choose a coordinate system with the z-axis anti-aligned with any temperature
gradient in the gas; and a spherical coordinate system with the pole ($\theta=0$) aligned with the z-axis.
For a sphere of radius $a$, Equation~(\ref{eq_photforce_general}) then reduces to
\begin{align}
{ F}_p& \cdot \ez = \nonumber\\ 
&-\frac{1}{2} \int_0^{2\pi} \int_0^\pi p\left( 2 +\frac{\alpha}{2} \left(\frac{T_s}{T_g}-1\right)  \right) a^2 \cos (\theta ) \sin(\theta) d\theta d\varphi.  
\label{eq_photforce_approx}
\end{align}
If the particle surface temperature $T_s$ is expressed as a Legendre series
\begin{align}
T_s(\theta) &= \sum_{n=0}^\infty A_n a^n P_n(\cos(\theta)),
\end{align}
then Equation~(\ref{eq_photforce_approx}) yields
\begin{align}
{ F}_p &= -\frac{\pi}{3} \alpha \frac{p}{T_g} a^3 A_1 \ez.
\label{eq_photoforce_a1}
\end{align}
This form for the photophoresis force on the sphere is commonly employed, 
 and a significant part of the literature deals with approximations for $A_1$ \citep{1993PhFl....5.2043B,rohatscheck1995,2012A&A...545A..36L}.

\subsection{Intensity Field in an Optically Thick Medium}
The frequency-dependent intensity field to linear order in an optically thick medium 
with the temperature varying along the z-axis is given by
\begin{align}
I_\nu (z,\mu) \approx B_\nu(T_g) - \frac{\mu}{\alpha_\nu+\sigma_\nu} \frac{\partial B_\nu(T_g) }{\partial T_g} \frac{\partial T_g}{\partial z} + \mathrm{O}(z^2),
\end{align}
where $\mu = \cos (\theta)$, $B_\nu(T_g)$ is the Planck function, and $\alpha_\nu$ and $\sigma_\nu$ are the absorption and scattering opacities  \citep{1986rpa..book.....R}.
Introducing the Rosseland Mean Opacity as
\begin{align}
\frac{1}{\kappa_R \rho_g} \equiv \frac{ \int_0^\infty (\alpha_\nu + \sigma_\nu)^{-1} \frac{\partial B_\nu}{\partial T} d\nu }{ \int_0^\infty \frac{\partial B_\nu}{\partial T} d\nu },
\end{align}
and integrating over frequency $\nu$, the frequency-integrated intensity field near $z=0$ is then given by
\begin{align}
I(\theta) &= \frac{\sigma_{SB}T_g^4 }{\pi}   \left[ 1 + 4\Gamma \cos \theta \right], \label{eq_Itheta} \\
\Gamma &\equiv -\frac{1}{\kappa_R \rho_g } \frac{\partial \ln T_g}{\partial z}.
\end{align}
This nondimensional parameter $\Gamma$ is therefore 
the important control parameter in the theory.
In an optically thick medium, one would expect that typically $\Gamma\lesssim 1$ as a steeper 
gradient would be rapidly smoothed by radiative thermal diffusion.

\subsection{Incident Intensity on a Small Sphere}

\begin{figure}
\begin{center}
\includegraphics[width=6cm]{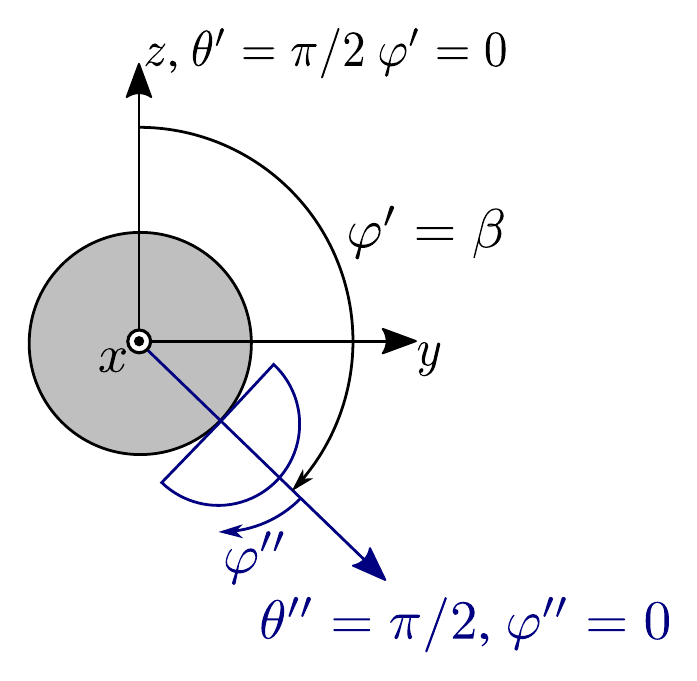}
\end{center}
\caption{Geometry for the integral of incident intensity over the surface of a sphere}
\label{fig_integral_geometry}
\end{figure}

{To determine the temperature structure of a sphere we need to integrate the incident intensity visible from any given
point of the sphere's surface. 
Here we make use the the assumption that the sphere is opaque, and that the sphere has a size significantly 
larger than the wavelength corresponding to the temperature of the optically thick radiation field.
Thanks to azimuthal symmetry around the $z$-axis for a sphere centered on the origin,
calculating the incident intensity on the circle defined by intersection of the sphere with the $yz$ plane suffices. Motivated
by the fact that the $x$-axis is perpendicular to that circle, we will recalculate the intensity (Equation~\ref{eq_Itheta}) in a 
spherical coordinate
system with the $x$-axis as its pole. Noting further that this intensity is unchanged by a spatial translation, we define additional coordinate
systems which are merely translations of the second coordinate system that place their origins on the circle. The 
angle between the surface and the radiation field, and appropriate limits
of integration for the incident intensity are then straightforward to determine.}

{As planned, we first} rewrite the intensity field in a new coordinate system, 
{ obtained by rotating the original one about the $y$ axis},
with the $\theta' = \pi/2$, $\varphi'=0$ axis directed  along $z$,
and the pole ($\theta'=0$) lying along the $x$ axis. In this coordinate system, Equation~(\ref{eq_Itheta}) becomes
\begin{align}
I(\theta',\varphi') = \frac{\sigma_{SB}}{\pi} T_g^4 [1 + 4 \Gamma  \sin \theta' \cos \varphi'].
\end{align}
Now translate this second coordinate system along $(\theta'=\pi/2, \varphi'=\beta)$ to the surface of the sphere,
 to form a third coordinate system with $\theta'' = \theta'$, and $\varphi''=0$ corresponding to $\varphi'=\beta$, requiring that $\beta \le \pi$.
 Note specifically that the poles of the second and third coordinate systems are aligned, and that the pole of the third coordinate
 system is tangent to the sphere.
As shown in Figure~\ref{fig_integral_geometry},
for $\theta''=\pi/2$, we have $\beta=\theta$ from the original coordinate system. 
{Because of the azimuthal symmetry around the radiation field,}
the incident flux on any point
on a sphere defined through $(\theta=\beta, \varphi \in [0,2\pi])$ in the original coordinate system
is the same as the incident flux on the {oriented} point defined through $(\theta''=\pi/2, \varphi''=0)$ in the third coordinate system.

We now integrate over the  solid angle visible from the point defined through $(\theta''=\pi/2, \varphi''=0)$.
The projection of a ray direction onto the surface normal ${\hat{n}} $ is $ \sin \theta''\cos \varphi''$, 
and we require the projection of each inwards going ray  {(for each visible direction $\beta +\varphi''$ 
the ray incoming along $\beta + \varphi''+\pi$)}
 onto the inward directed surface normal $-{\hat{n}}$.
{Further, by construction both the pole $(\theta''=0)$ and the vector
$(\theta''=\pi/2, \phi''=\pi/2)$ are tangent to the sphere's surface, which sets the limits of integration.}
 The integral for the
incident intensity on the surface of a small sphere is therefore
\begin{align}
I_n&( \beta) = \int_\mathrm{outward} I(\theta'',\varphi'' )\cdot ({-\hat{n}}) \, d\Omega \\
=&  \int_0^\pi \int_{-\pi/2}^{\pi/2} I(\theta'',\beta + \varphi''+\pi)\sin \theta''  \cos \varphi''  \times \sin \theta'' d\varphi'' d\theta''\\
&= \sigma_{SB}T_g^4 \left[ 1 -\frac{8}{3}\Gamma \cos(\beta) \right].
\end{align}
This intensity is then the forcing on the boundary for the heat transfer calculation in the sphere, which  will yield the 
variation of the particle surface temperature $T_s$.

\subsection{Surface Temperature of an Irradiated Sphere}
To solve the problem of the heat transfer within a spherical particle we make several assumptions.
First, we include both thermal conduction to the gas, and thermal radiation from the particle surface.
The conductive heat flux away from the surface in a dilute gas given by \cite{1996JVST...14..588S}:
\begin{align}
I_c
& = \Upsilon T_g^{1/2} (T_g-T_s),  \\
\Upsilon &\equiv  \frac{\gamma +1}{8(\gamma -1)} \alpha   \sqrt{\frac{2 }{\pi}} \left(\frac{ k_B}{ \mu m_H}\right)^{3/2} \rho_g.
\end{align}
Second, we shall assume that the particle is opaque such that the irradiation imparts energy only 
on the spherical surface of the particle, and not inside the volume.
This breaks down for small, low volume filling fraction particles which might only have one or two
monomers along a given ray through the particle.

Under these assumptions the heat equation within the sphere is simply Laplace's equation $\nabla ^2 T= 0$,
and the solution can can be expressed as a series 
in $r^\ell P_\ell(\cos \theta)$ where the $P_\ell$ are Legendre polynomials.
The coefficients of the solution are obtained by solving the boundary condition equating the flux of energy
conducted from the surface and radiated from the surface to the incident energy:
\begin{align}
k\left. \frac{\partial T(r,\theta)}{\partial r}\right|_{a}  +\sigma_{SB}T(a,\theta)^4 -I_c(\theta) = I_n(\theta), \label{eq_solid_bc}
\end{align}
where $T(r,\theta)$ is the internal temperature of the particle, $k$ its thermal conductivity and $a$ is its radius. 
We decompose the particle temperature into a constant component and a spatially varying one:
\begin{align}
T(r,\theta)&= T_0+T'(r,\theta). \label{T_expansion}
\end{align}
We linearize the problem with the small parameter being the deviation of the particle temperature
\begin{align}
\eta = T'(r,\theta)/T_0 &\ll 1.
\end{align}
The linearized problem is then
\begin{align}
k \left.\frac{\partial T' }{\partial r}\right|_{a}  &+ \sigma_{SB} T_0^4 \left( 1+\frac{T'}{T_0}\right)^4 \nonumber\\
&   - \Upsilon T_g^{1/2} \left(T_g-T_0\left(1+\frac{T'}{T_0}\right)\right) \nonumber\\
&= \sigma_{SB} T_g^4 \left[ 1 -\frac{8}{3}\Gamma \cos(\theta) \right] .
\end{align}
By expanding the second term on the left hand side to linear order in $\eta$ and substituting a Legendre series for $T$ 
we obtain the solution in terms of the first two terms of the Legendre series:
\begin{align}
T_0 &=T_g \label{T0}
\end{align}
and
\begin{align}
&T'(r,\theta) = A_1r \cos(\theta) + \mathrm{O}\left[\left(A_1a\right)^2\right], \label{T'} \\
&A_1 = -\frac{8}{3}  \frac{ \sigma_{SB} T_g^4 \Gamma}{k} \left( 1 +4 \sigma_{SB} T_g^3  \frac{a}{k}  + \Upsilon T_g^{1/2}  \frac{a}{k} \right)^{-1}.
\label{linear_single_A1}
\end{align}
In Appendix~\ref{app_numericalsolution} we compare this to a solution of the full problem to verify 
the validity of the linear approximation in the protoplanetary disk regime.

Combining Equations~(\ref{zeta_definition}), (\ref{T_expansion}), (\ref{T0}), (\ref{T'}) and (\ref{linear_single_A1}) we find
\begin{align}
\zeta= -\frac{8}{3}  \alpha  \sigma_{SB} T_g^3 \Gamma\frac{a}{k} \left( 1 +4 \sigma_{SB} T_g^3  \frac{a}{k}  + \Upsilon T_g^{1/2}  \frac{a}{k} \right)^{-1}\cos(\theta).
\end{align}
For the  $\zeta \ll 1$ expansion of the force integral to hold to linear order,
we need one or a combination of $\Gamma \ll 1 $, $\Upsilon T_g^{1/2}  \frac{a}{k}  \gg 1$, or $\sigma_{SB} T_g^3  \frac{a}{k} \ll1$.

At this point we can insert this result into Equation~(\ref{eq_photoforce_a1})
to yield the photophoresis force as
\begin{align}
{ F}_p \approx  \frac{8\pi}{9} & \frac{ k_B }{\mu m_H}  \alpha \frac{a^3}{k}   \rho_g \sigma_{SB} T_g^4 \Gamma \nonumber\\
    &  \left( 1  +4 \sigma_{SB} T_g^3  \frac{a}{k}  + \Upsilon T_g^{1/2}  \frac{a}{k} \right)^{-1}  \ez \, .
  \label{eq_sphere_force}
\end{align}
where we have used $p= k_B \rho_g T_g/(\mu m_H)$.
We proceed to examine several limiting cases where the form of the force can be usefully simplified.

\subsubsection{Limits }
As mentioned in the previous section, the validity of the $\zeta \ll 1$ expansion of the force integral
requires one or a combination of  $\Gamma \ll 1 $, $\Upsilon T_g^{1/2}  \frac{a}{k}  \gg 1$, or $\sigma_{SB} T_g^3  \frac{a}{k} \ll1$.
The first case does not lead to a significant simplification, and the second 
case leads to a very small photophoresis force.
In the third case the photophoresis force on a single-layered spherical particle simplifies to
\begin{align}
{ F}_p &\approx  \frac{8\pi}{9}  \frac{ k_B }{\mu m_H}  \alpha \frac{a^3}{k}   \rho_g \sigma_{SB} T_g^4 \Gamma  \left( 1  + \Upsilon T_g^{1/2}  \frac{a}{k} \right)^{-1}  \ez.
  \label{eq_sphere_force_zeta}
\end{align}
Furthermore, where the heat transfer by conduction from the particle to the gas is much smaller than the
radiative cooling of the particle, the force further simplifies:
\begin{align}
{ F}_p &\approx  \frac{8\pi}{9}  \frac{ k_B }{\mu m_H}  \alpha \frac{a^3}{k}   \rho_g \sigma_{SB} T_g^4 \Gamma  \ez.
  \label{eq_sphere_force_zeta_rad}
\end{align}
Equation~(\ref{eq_sphere_force_zeta_rad}) gives a force which is always greater than the more general form
in Equation~(\ref{eq_sphere_force}). In Section~\ref{sec_heat_transfer_negligible} we will determine where this difference becomes significant
for a specific particle model.

\subsection{ Surface Temperature of an Trradiated Double-layer Sphere}
\begin{figure}
\plotone{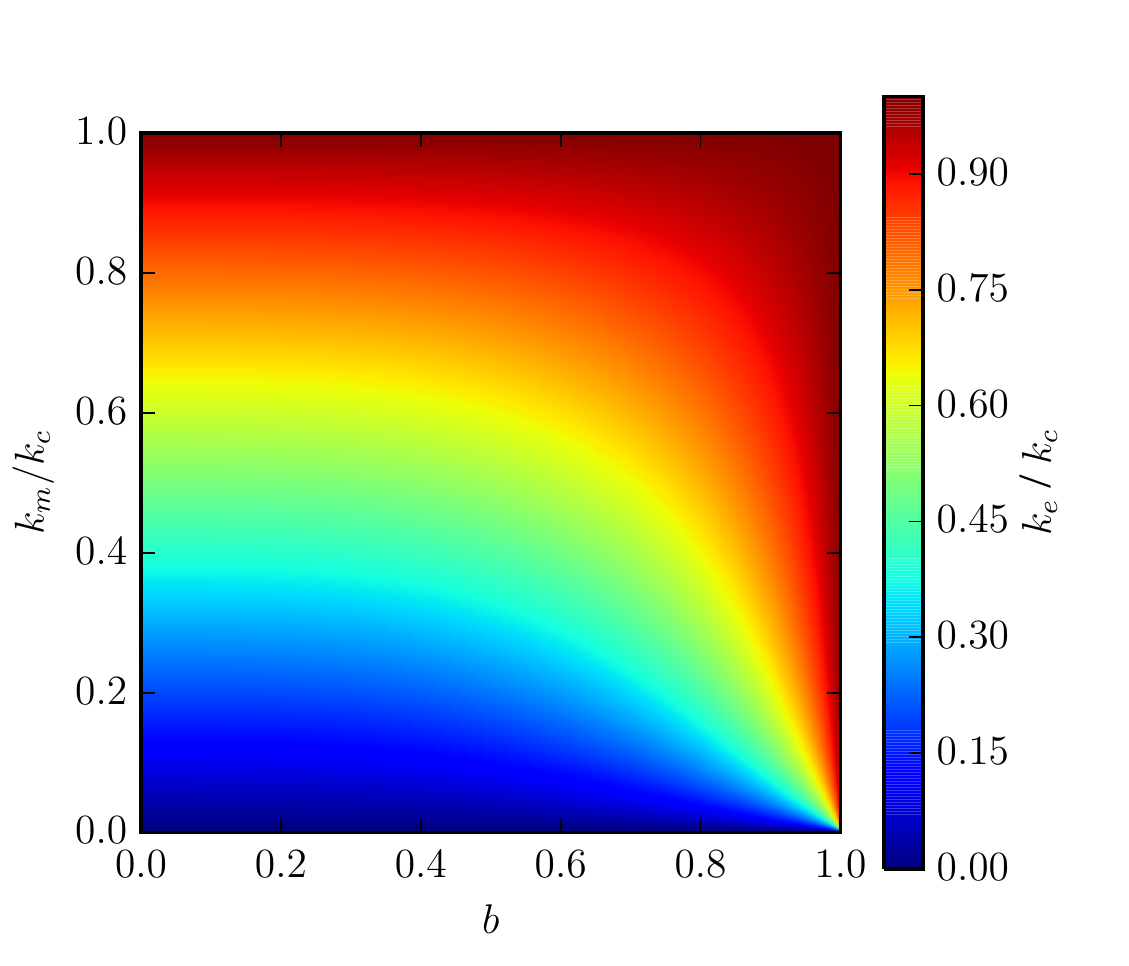}
\caption{Equivalent conductivity $k_e$ (Equation~\ref{eq_k_e}) of a double-layer sphere in terms of a single-layer 
sphere when radiation dominates the particle cooling
and effective $\sigma_{SB} T_g^3  \frac{a}{k} \ll1$.}
\label{fig_equiv_conduct}
\end{figure}

As particles with outer rims of a material with differing thermal conductivity are of interest in protoplanetary disks
we proceed now to develop a similar approximation for double layered particles.
For a double-layer sphere with the same basic assumptions as for the single layer sphere,
 we develop the solution in terms of series expansions in the core and mantle $\left(T_c(r,\theta)\right.$ and $\left.T_m(r,\theta)\right)$ layers as
\begin{align}
T_c(r,\theta) &= \sum_{n=0}^\infty \left( C_n r^n  \right) P_n(\cos(\theta)), \\
T_m(r,\theta) &= \sum_{n=0}^\infty \left( A_n r^n +  B_n r^{-n-1} \right) P_n(\cos(\theta)).
\end{align}
The boundary conditions at the core radius $r_c$ match the energy flux and temperature 
between the core with thermal conductivity $k_c$ and mantle with thermal conductivity $k_m$ through
\begin{align}
k_c \left. \frac{\partial T_c}{\partial r} \right|_{r_c} &= k_m \left. \frac{\partial T_m}{\partial r}\right|_{r_c}, \\
T_c(r_c,\theta) &= T_m(r_c,\theta),
\end{align}
where the second equation assumes that the interface has perfect conduction.
At the exterior, mantle plus core radius $a$, the boundary condition is  again Equation~(\ref{eq_solid_bc})
although with the mantle's thermal conductivity $k_m$.

The linear approximation in $\eta\ll1$ for this problem and its 
solution are given in Appendix~\ref{app_lineardoublesolution}. The result for the
  photophoresis force on a double-layered sphere in a dilute optically thick medium is:
\begin{align}
 { F}_p  \approx &
   \frac{8\pi }{9}   \frac{ k_B }{\mu m_H}  \alpha   \frac{a^3}{k_m} \rho_g \sigma_{SB}T_g^4 \Gamma 
  \left[(1-b^3)+\frac{k_m}{k_c} (2+b^3)  \right] \nonumber\\
&  \times \left[
    2  \frac{k_m}{k_c} (1-b^3) + \left( 1 +2 b^3  \right)
     + \frac{a}{k_m} \left(  (1 - b^3)   \right. \right. 
     \nonumber\\
  &  \quad \left. \left. 
  +  \frac{k_m}{k_c} (2+b^3) \right) \left( \Upsilon T_g^{1/2}  +4\sigma_{SB}T_g^3\right)
    \right]^{-1}  \ez   ,
    \label{eq_doublesphere_force}
\end{align}
where $b=r_c/a$ is the radius of the core in terms of the exterior radius of the sphere.

Comparing Equation~(\ref{eq_doublesphere_force}) with Equation~(\ref{eq_sphere_force})
in the case that radiative cooling dominating over conductive cooling, 
requiring additionally $\sigma_{SB} T_g^3  \frac{a}{k_m} \ll1$,
gives the equivalent single-layer conductivity of a double-layered sphere with the same
meaning as given for the optically thin case by  \citet{2012A&A...545A..36L}:
\begin{align}
k_e = &{k_m} 
  \left[(1-b^3)+\frac{k_m}{k_c} (2+b^3)  \right] ^{-1} \nonumber\\
&  \times \left[
    2  \frac{k_m}{k_c} (1-b^3) + \left( 1 +2 b^3  \right)
    \right]. \label{eq_k_e}
\end{align}
This is illustrated in Figure~\ref{fig_equiv_conduct}, showing that 
mantle layers of low conductivity material do dramatically alter
the effective property of a particle with $b \lesssim 5/6$.

\subsection{Photophoresis Drift Velocity}

The drag force $\bm{F}_d$ is the aerodynamic drag on particles, and can be written as
\begin{equation}
 \bm{F}_d = -m_d \frac{\bm{v}}{\tau}, \label{eq_drag}
 \end{equation}
 where $\tau$ is the particle frictional stopping time. It will be useful to define the particle Stokes number
 \begin{equation}
 St \equiv \tau \Omega_K \label{eq_stokesdef}
 \end{equation}
 where $\Omega_K$ is the local Keplerian frequency .

In practice, small, sub-centimeter particles in protoplanetary disks are in the `free molecular flow' regime, i.e. the Epstien drag regime. 
From \citet{1996Icar..124..441B}'s experiments on the Brownian motion diffusion of particles, under these conditions
Equation~(\ref{eq_drag}) takes the form
\begin{align}
\bm{F}_d = -\frac{\sigma_a \rho_g v_g}{\epsilon} \bm{v}, \label{drag_0}
\end{align}
with $\epsilon = 0.68$ an experimentally determined constant,
$v_g$ the mean thermal speed of gas particles
\begin{align}
v_g = \sqrt{\frac{8 k_B T_g}{ \pi  \mu m_H}},
\end{align}
and $\sigma_a=\pi a^2$ the geometric cross section of the particle.

Balancing the drag force and the photophoresis force in the approximation of 
Equation~(\ref{eq_sphere_force}) gives a drift speed of 
\begin{align}
v_p =  \frac{1}{9}&\sqrt{\frac{8 \pi k_B}{\mu m_H} } \, \epsilon \alpha   \frac{a}{k} \sigma_{SB} T_g^{7/2} \Gamma  \nonumber\\
     &\times\left( 1  +4 \sigma_{SB} T_g^3  \frac{a}{k}  + \Upsilon T_g^{1/2}  \frac{a}{k} \right)^{-1}.
     \label{eq_vp}
\end{align}

\subsection{When does photophoresis dominate over radiation pressure?}

In SI units the force due to radiation pressure on a black sphere with radius $a$ is:
\begin{align}
{ F}_r &= \frac{{I}}{c} \pi a^2,
\end{align}
and here the Rosseland mean flux can be used for $ I$, as the relevant quantity is the net momentum of the photons impacting the particle.
Following \cite{1986rpa..book.....R} we write
\begin{align}
{I} &=  \frac{16 }{3 } \sigma_{SB}  T_\mathrm{g}^4 \Gamma \ez.
\end{align}
Treating the drag force in the same manner as with photophoresis, the resulting drift speed of the particle is
\begin{align}
v_r = \frac{18}{3} \sqrt{\frac{\pi \mu m_H}{8 k_B} }  \frac{\epsilon}{c \rho_g} \sigma_{SB} T_g^{8/2} \Gamma.
\label{eq_rp_drift}
\end{align}
Due to the linear nature of the Epstien drag regime, the drift velocities due to the two forces simply add linearly.
Accordingly, the ratio between the drift speeds is 
\begin{align}
\frac{v_p}{v_r} = \frac{1}{54} \frac{8 k_B}{\mu m_H} c \rho_g \alpha \frac{a}{k} T_g^{-1}  \left( 1  +4 \sigma_{SB} T_g^3  \frac{a}{k}  + \Upsilon T_g^{1/2}  \frac{a}{k} \right)^{-1},
\end{align}
where Equation~(\ref{eq_sphere_force}) has been used for the photophoresis force to ensure generality across parameter space.
Radiation pressure dominates at low density, high temperature and high conductivity.
These parameters fall mainly outside the regime of the optically thick part of a protoplanetary disk. \
However, if desired, the combined drift velocity of dust due to photophoresis and radiation pressure in a dilute optically thick 
medium can be obtained from the formulas already given in this work.

\subsection{When can heat transfer by conduction be neglected?}

\label{sec_heat_transfer_negligible}
\begin{figure}
\begin{center}
\plotone{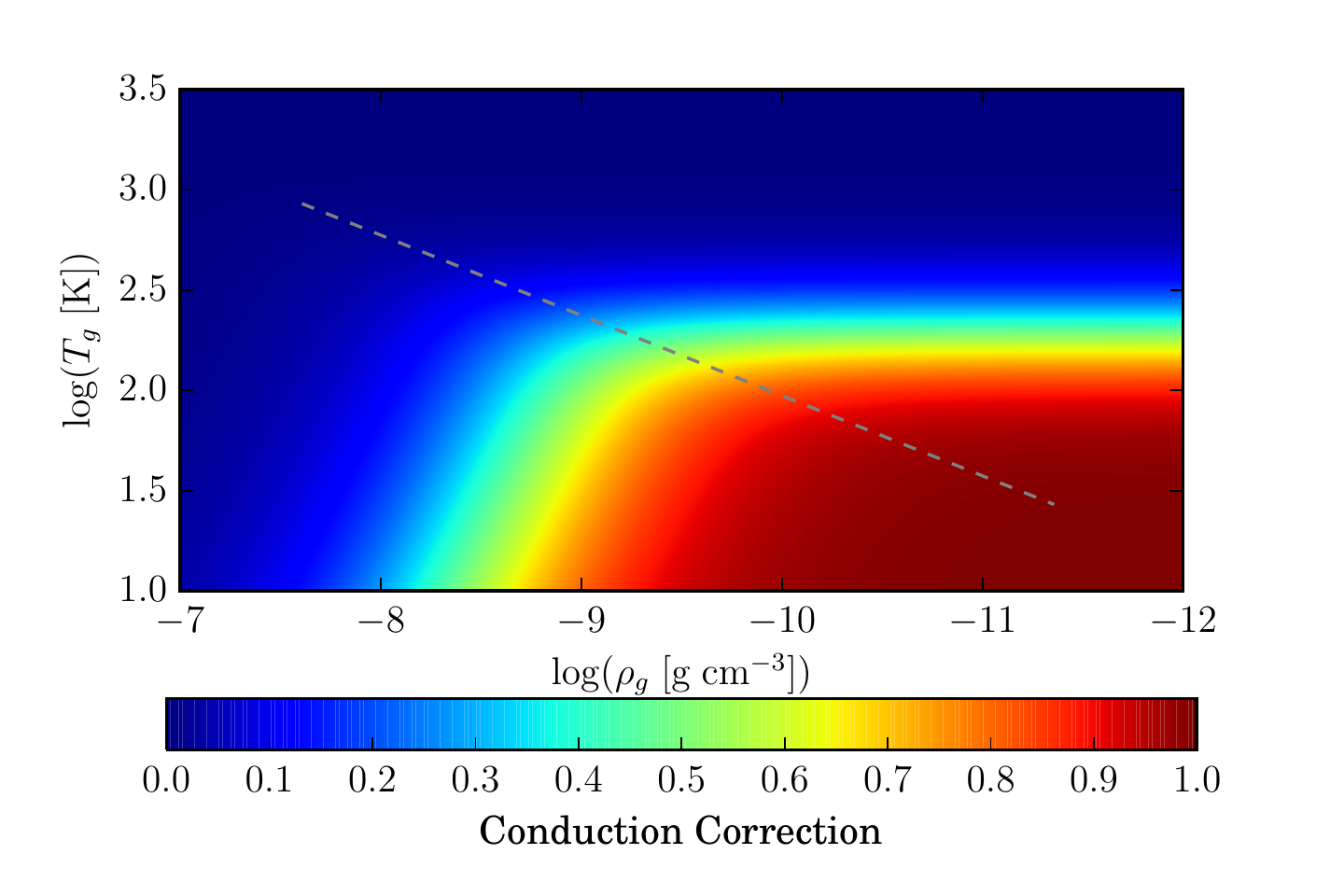}
\caption{Conduction correction for a very porous $1\ \mathrm{mm}$ silicate aggregate
 particle with accordingly very low thermal conductivity $1.5\times10^2\ \mathrm{erg\ s^{-1} cm^{-1} K^{-1}}$.
The dashed curve is the MMSN midplane density and temperature. }
\label{figcondcorr}
\end{center}
\end{figure}
The question at hand is when does Equation~(\ref{eq_sphere_force_zeta_rad}) provide a good approximation 
to Equation~(\ref{eq_sphere_force})? The latter always yields a  smaller force, so when the 
approximation ceases to hold the photophoresis force is effectively shut off.
For the sake of this discussion, we refer to the factor
\begin{align}
\left( 1  +4 \sigma_{SB} T_g^3  \frac{a}{k}  + \Upsilon T_g^{1/2}  \frac{a}{k} \right)^{-1}
\end{align}
in the photophoretic force for a sphere as given by Equation~(\ref{eq_sphere_force})
as the `conduction correction'.
Figure~\ref{figcondcorr} gives this conduction correction
for a  $1\ \mathrm{mm}$ particle with conductivity $1.5\times10^2\ \mathrm{erg\ s^{-1} cm^{-1} K^{-1}}$. 
The iso-countours of this plot are roughly two power-laws, breaking coincidentally around the
 MMSN midplane density-temperature relation.
 It can be roughly said that below the break the contour is set by the effect of heat conduction to the surrounding gas ($\Upsilon T_g^{1/2}a/k$), 
 while above the break it is set by the $4\sigma_{SB}T_g^3a/k$ term. 

\begin{figure}
\begin{center}
\plotone{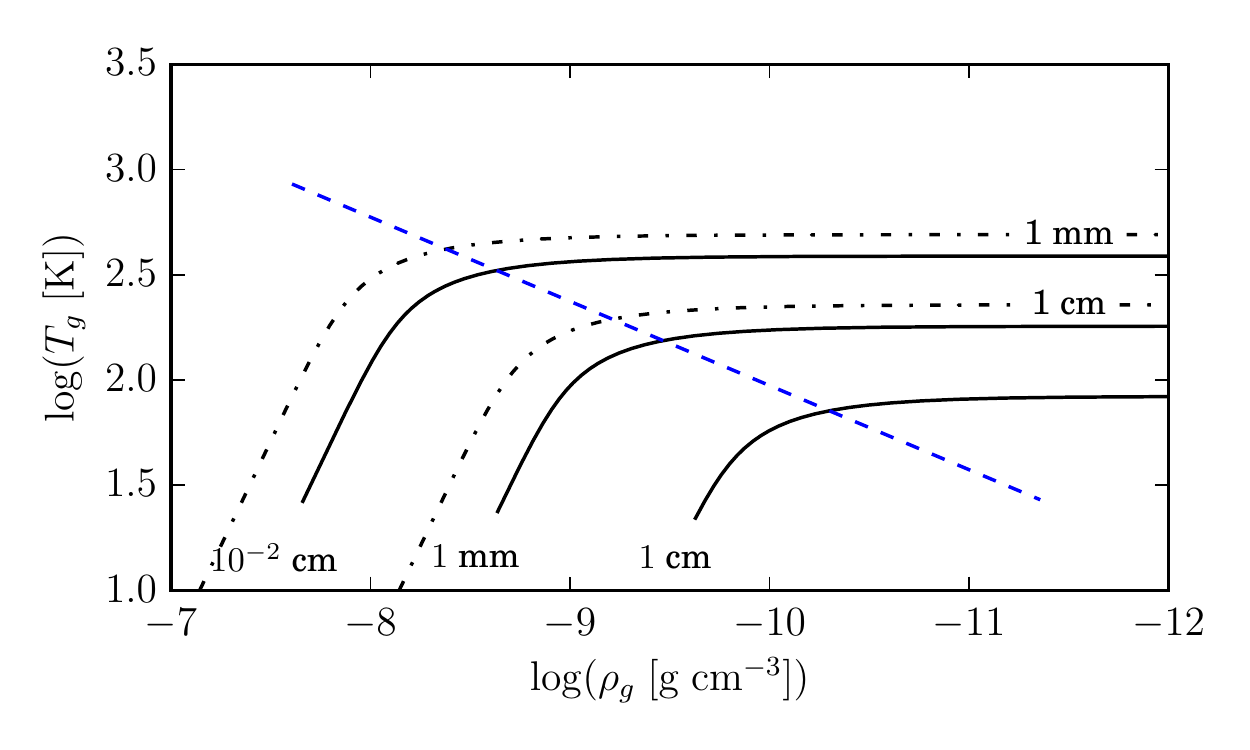}
\caption{Contours where the extra conduction terms decrease the photophoresis force by a factor of $50\%$.
Contours are are labelled with particle radius $a$, and to the left and above of each contour conduction has reduced the 
photophoresis force by more than half. Solid lines are for particles with volume filling factor $\phi=0.12$ 
and dash-dot lines are for particles with $\phi=0.5$.
 }
\label{figcriticala}
\end{center}
\end{figure}

In Figure~\ref{figcriticala} a more general result is plotted, with curves of the critical contour in $T_g$-$\rho_g$ space
where the difference between Equation~(\ref{eq_sphere_force_zeta_rad}) and Equation~(\ref{eq_sphere_force})
is $50\%$. For large, porous particles, which otherwise have the largest photophoresis forces the
 conduction terms have the most dramatic effect in shutting down photophoresis.

\section{Applications in Protoplanetary Disks}

\label{sec_applications}

\subsection{Particle Model}
For the purposes of estimating the effects of photophoresis in a protoplanetary disk, we adopt a model of particles following
from the experimental measurements of the thermal conductivity of packed silicate monomers by \citet{2011Icar..214..286K}.
These give well constrained values of the thermal conductivity so vital to determining the photophoresis force,
finding
\begin{equation}
k(\phi) = k_0 e^{7.91\phi} ,
\label{k_of_phi}
\end{equation}
with 
\begin{equation}
k_0=5.14\times10^{-4}\ \mathrm{W\ m^{-1}K^{-1}},
\end{equation}
or $k_0=51.4\ \mathrm{erg\ s^{-1} cm^{-1} K^{-1}}$.
In Equation~(\ref{k_of_phi}), $\phi$ is the grain's volume filling factor, { defined for a given sample as
$\phi\equiv \rho_d/\rho_0$ where $\rho_d$ is the density of the bulk sample and $\rho_0$ the 
mass density of the constituent solid silicate monomers \citep{2011Icar..214..286K}. }
Other grain materials, such as organics/graphite or mixed material grains are of
great interest in planet formation, but such experimental data does not yet exist for those materials,
so we only consider silicate particles,
 and vary the volume filling factor and radius of the particles.

For our model aggregates, a dust grain of mass $m_d$ 
has a density and radius:
\begin{align}
&\rho_d = \phi \times \rho_0, \\
&a = \left(\frac{3m_d}{4\pi \rho_0}\right)^{1/3} \phi^{-1/3}.
\end{align}
where $\rho_0 \simeq 3\times 10^3 \ \mathrm{kg\,m^{-3}}$ is the density of fused silicates of chondritic composition.

We assume for simplicity that non-thermally processed dust grains are composed of
\begin{equation}
a_0 = 0.75\ \mu \mathrm{m}
\end{equation}
 sized monomers \citep{2010A&A...513A..56G}.
For silicate dust, the results of \citet{2015Icar..254...56H} imply that thermal processing for micron sized grains occurs on several week timescales at $T=900$\,K and several
decade timescales at $T=850$\,K, and so for $T<850$\,K, a fluffy porous structure is expected, while for $T>900$\,K, the dust grains should be fused.
We further normalize our grain masses to
\begin{equation}
m_{d0} = 2 \times 10^{-4}\ \mathrm{g} \simeq \frac{4\pi}{3} \rho_0 \times \left(0.25 \ \mathrm{mm}\right)^3,
\end{equation}
the mass corresponding to a typical chondrule found in the  LL3.0 chondrite Semarkona \citep{2014arXiv1408.6581F}.
In the application here, we consider that the particle has the same mean temperature as the gas at its location. 
For mm-size and smaller paticles in a protoplanetary disk, this is a reasonable assumption for timescales longer than minutes \citep{paper1}.

\subsection{MMSN Model}
To examine particle motion in a protoplanetary disk we adopt the Hayashi minimum mass solar nebula \citep[MMSN,][]{1981PThPS..70...35H}
as our canonical disk model, with its temperature set by the stellar irradiation:
\begin{align}
T_g &= 270 R^{-1/2}\ \mathrm{K}, \\ 
\Omega_K &= 2 \times 10^{-7} R^{-3/2}\ \mathrm{s}^{-1}, \\
H &= \sqrt{\frac{p}{\rho_g}} \Omega_K^{-1}, \label{H_equation} \\
\rho_g &= \rho_{g0} e^{-z^2/2H^2}, \label{rho_g_equation} \\
\Sigma_g &= \sqrt{2\pi} \rho_{g0} H = 1700 R^{-3/2} \ \mathrm{g\ cm}^{-2}. \label{Sigma_g_equation}
\end{align}
$T_g$ is the background disk temperature,
{  $\Omega_K$ is the Keplerian orbital angular velocity,
$H$ is the scale height assuming vertical hydrostatic equilibrium,
$\rho_{g0}$ is the midplane gas volume density,
 $\Sigma_g$ is the disk gas surface density, }
 and $R$ the orbital position measured in AU. 
We can rewrite Equation~(\ref{H_equation}) as
\begin{equation}
\Omega_K^2 = \frac{p}{\rho_g} H^{-2} = \frac{k_B T_g}{\mu m_H} H^{-2} \label{Omega_Eq}.
\end{equation}
This provides the background in which we examine optically thick photophoresis.
We follow \citet{paper1} in assuming that the opacity in the medium is provided by small grains 
which can be assumed to be well-coupled to the temperature of the gas.

\subsubsection{Temperature Gradients}

\citet{2014ApJ...791...62M} is the only study of self-consistent 
heating and cooling of current sheets in a local simulation of a protoplanetary disk flow 
we are aware of at this time which allows the 
identification of maximal temperature variations and temperature gradients in the optically thick regime.
Analyzing those simulations, we find that they contain well resolved temperature 
gradients with maximum $\Gamma$ varying typically between $1-7\times 10^{-4}$ over time.
Hence, we take that result as 
an estimate for local temperature fluctuations in
 the analysis of protoplanetary disks, normalizing our results to a
 fiducial value of $\Gamma=10^{-4}$.
We caution that $\Gamma$ is not known to be universal, and
other disk flow regimes may have different typical maximal values of $\Gamma$. 
We hope that this work will motivate further studies characterizing other regimes.

\subsection{Particle Drift Speeds are Very Porosity Dependent}
\begin{figure}
\begin{center}
\plotone{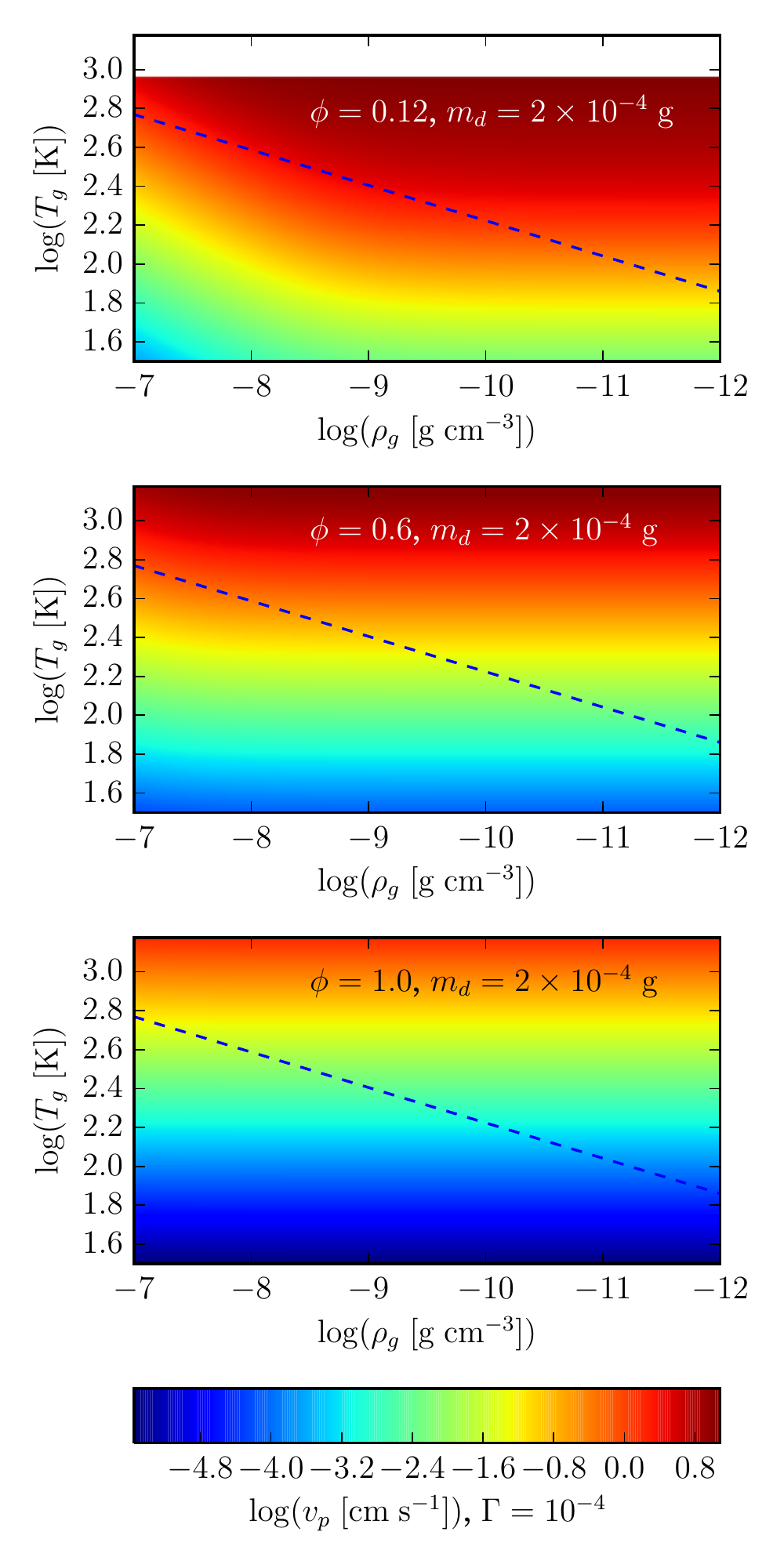}
\caption{
Photophoresis driven drift velocities (Equation~(\ref{eq_vp}) for particles of a fixed mass and varying 
porosity driven by a fluctuation with temperature gradient $\Gamma=10^{-4}$.
When solid, with $\phi=1$, these particle have a radius of $1\ \mathrm{mm}$.
The dashed line gives the MMSN midplane density-temperature relation.
In the upper panel, no data is shown above $T_g = 900\ \mathrm{K}$ as silicate grains
will rapidly compact at temperature similar to this, leaving no very porous grains at higher temperatures.
}
\label{figvp}
\end{center}
\end{figure}

Figure~\ref{figvp} shows photophoresis driven particle drift speeds for particles with mass $2 \times 10^{-4}\ \mathrm{g}$ and varying porosity.
For these equal mass particles, the drift speeds vary greatly depending on the porosity.
With our fiducial $\Gamma=10^{-4}$ temperature gradient 
the speed differences are commonly an order of magnitude at the MMSN midplane and higher temperatures.

\subsection{Comparison to Radial Drift Speeds}

\begin{figure}
\begin{center}
\plotone{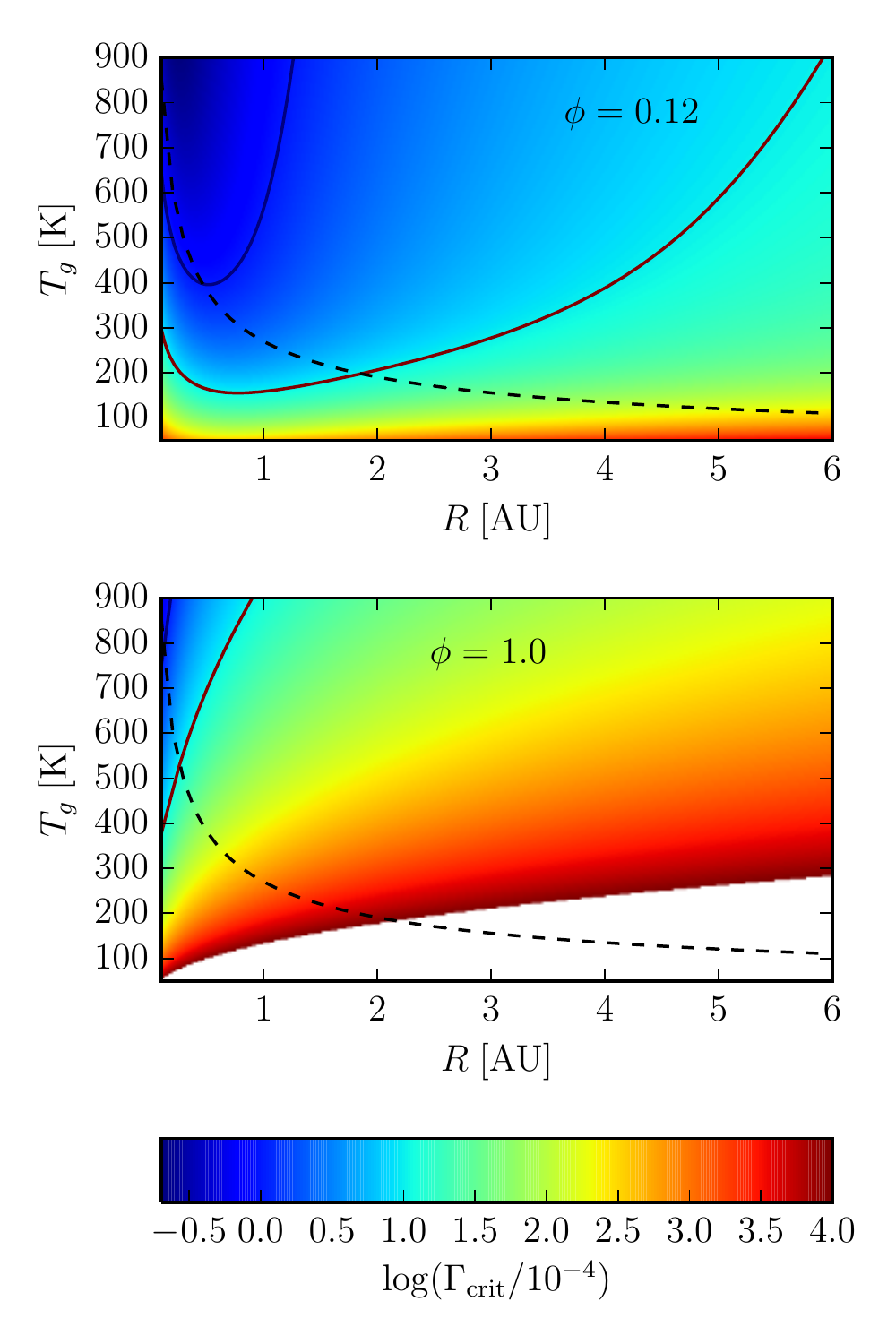}
\caption{
Critical temperature gradient $\Gamma_\mathrm{crit}$ for 
the photophoresis driven drift velocity to equal the radial drift velocity.
Solid contours are drawn at $\Gamma_\mathrm{crit}=10^{-4},10^{-3}$, and the values are cut off at $\Gamma_\mathrm{crit}>1.0$.
The dashed line shows the MMSN midplane temperature radius relation.
{ This indicates that at small radii photophoresis driven collisions can dominate 
over radial drift driven collisions when $\Gamma \gtrsim 10^{-4}$, particularly for porous grains.}
}
\label{figgammacrit}
\end{center}
\end{figure}

The photophoretic drift speed $v_p$ depends on the particle's size and thermal conductivity (and hence its porosity). 
Two particles with different $(a,\phi)$
would therefore experience different drift speeds $v_p(a,\phi)$ even in the same gas environment. 
This results in a relative drift which can drive
collisions with a speed comparable to the higher photophoretic drift speed of the two particles. 
We can check the relevance of photophoresis due
 to temperature fluctuations on particle
collisions by comparing $v_p$ to other dust drift speeds.
 The most common drift speed considered for collisions is dust settling, which is complicated in the presence
of photophoresis due to photophoretic levitation (see Section~\ref{sec_levitation}).
 However dust also drifts radially.

This radial drift is due to the inwards pointing radial pressure gradient in protoplanetary disks, which causes the gas to orbit at a sub-Keplerian velocity. 
Inertial particles do not feel the pressure gradient and so orbit faster than the gas and feel a headwind which slows them, and causes them to gradual
inspiral.
Following \citet{1977MNRAS.180...57W} we write
the orbital velocity of the gas as
\begin{equation}
u_g = (1-\eta) u_K,
\end{equation}
where
\begin{equation}
u_K = R \Omega_K
\end{equation}
 is the local Keplerian velocity and $\Omega_K$ the Keplerian frequency.
In an MMSN with $T\propto R^{-1/2}$, we have $\eta \ll 1$, and $\eta u_K \simeq 45\ \mathrm{m\ s^{-1}}$, independent of the orbital position. 
We can write the radial velocity equation for particles well coupled to the gas (i.e.~the particle's orbital velocity matches the gas) under the assumption
that terminal radial velocity has been achieved (Equation~\ref{eq_drag}):
\begin{align}
\frac{\partial v_r}{\partial t} = -\frac{v_r}{\tau} + \frac{v_{orb}^2}{R} - \frac{u_K^2}{R} = 0.
\end{align}
Using Equation~(\ref{eq_stokesdef}) we can write the equation for the headwind induced inspiral drift speed
\begin{align}
v_r \simeq -2 St\, \eta\, u_K. \label{v_radial_drift}
\end{align}
Like photophoresis, this will drive collisions between particles with different drag coefficients.

Photophoresis from local temperature fluctuations
will drive more powerful and
frequent collisions if $|v_p|>|v_r|$, i.e. if
\begin{equation} 
\frac{\tau F_p}{m_d} > 2 St \eta u_K.
\end{equation}
From this, we can determine the critical value $\Gamma_\mathrm{crit}$ where photophoresis driven velocities equal
 the radial drift velocity at the midplane
\begin{align}
\Gamma_\mathrm{crit} =  3 & \sqrt{\frac{2 \pi \mu m_H}{ k_B }}   \frac{k}{\alpha \sigma_{SB} }   \frac{  \rho_0} {\Sigma_g}    T_g^{-7/2} \nonumber\\
    &  \left( 1  +4 \sigma_{SB} T_g^3  \frac{a}{k}  + \Upsilon T_g^{1/2}  \frac{a}{k} \right)  \eta u_K 
\end{align}

Figure~\ref{figgammacrit} gives the size of temperature gradient due to local temperature fluctuations
 needed for  photophoresis driven drift to equal the MMSN radial drift speed ($\eta u_K=45\mathrm{\ m\ s^{-1}}$) for particles with 
small Stokes number. These are normalized to our fiducial value of $\Gamma=10^{-4}$.
{ In an MMSN, porous ($\phi=0.12$) grains satisfy $\Gamma_\mathrm{crit}<10^{-3}$ inwards of $2$\,AU, and
 $\Gamma_\mathrm{crit}<10^{-4}$ inwards of $0.5$\,AU. Future studies of the appropriate parameter
 range for $\Gamma$ are therefore needed for studies of collisional grain growth in the inner few AU of protoplanetary disks.}

\subsection{Dust Levitation in a MMSN Protoplanetary Disk}
\label{sec_levitation}

\begin{figure*}
\begin{center}
\plotone{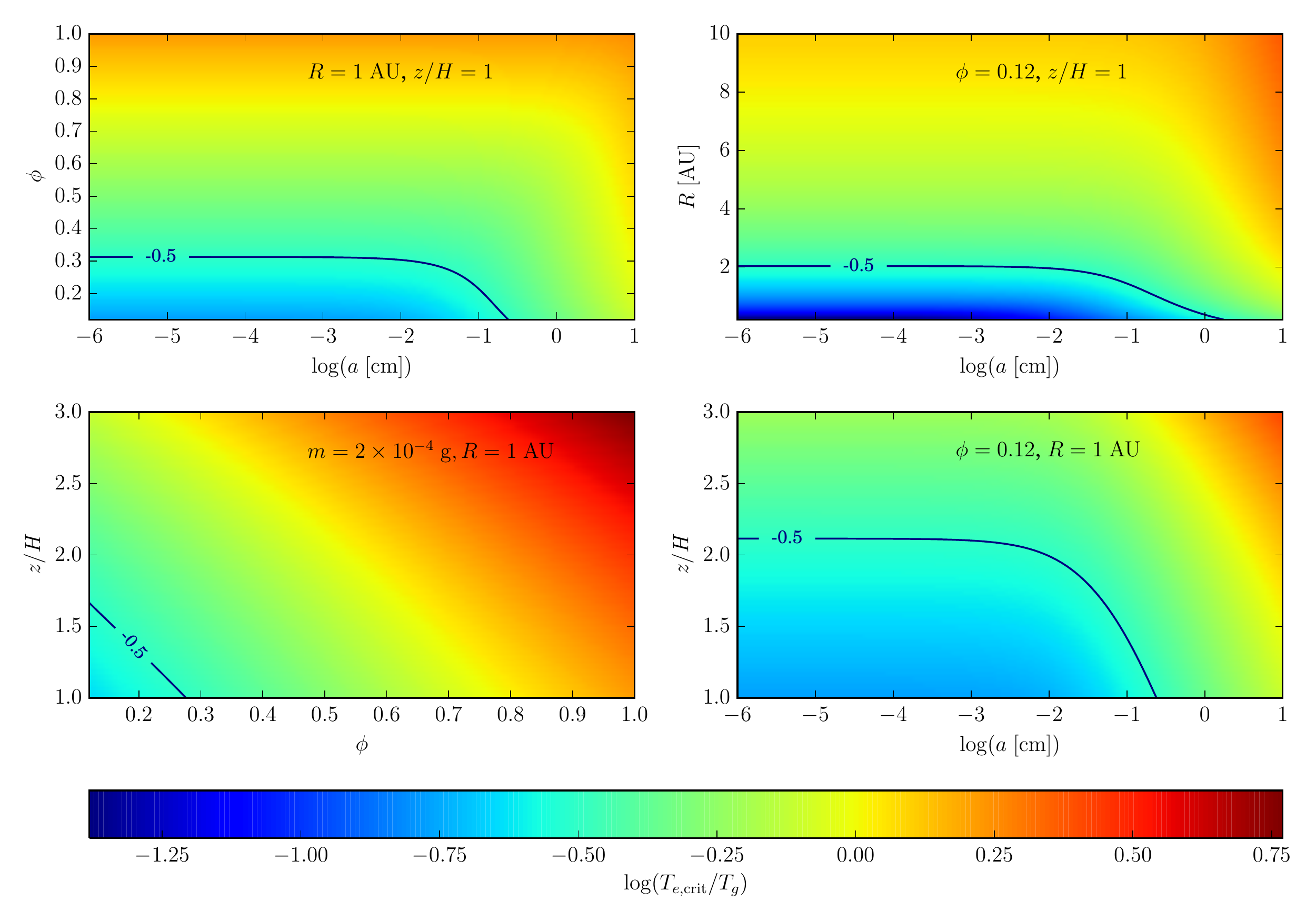}
\caption{
Critical effective temperature due to accretion heating $T_\mathrm{e,crit}$ which will produce dust levitation for various parameters.
Plotted is the ratio $T_\mathrm{e,crit}/T_g$, where $T_g$ is the actual local gas temperature, and
the solid contours show an example level $\log(T_\mathrm{e,crit}/T_g)=-0.5$.
Where is this condition is fulfilled 
by the disk structure the region below the contour would correspond to dust photophoretically levitated in the disk.
{\it Upper Left:} as a function of radius $a$ and porosity $\phi$ at $1\ \mathrm{AU}$ and at the height of one scale height in a MMSN.
At this location, levitation cuts off for particles above $a \sim 1\ \mathrm{mm}$.
{\it Upper Right:} as a function of particle radius $a$ and radial position $R$ for a fixed very low porosity $\phi$ at one scale height.
We see that at small radial positions levitation may occur, but would result in only particles with $a \lesssim 1\ \mathrm{mm}$ being levitated.
{\it Lower Left:} For particles with the mass of a typical chondrule at $1\ \mathrm{AU}$ photophoresis has a weak ability to levitate, 
with the only possibility being for the most porous particles.
{\it Lower Right:} At $1\ \mathrm{AU}$ small, porous particles may be levitated to several scale heights.
}
\label{figlevitation}
\end{center}
\end{figure*}

Here we find that levitation is also possible in the optically thick region of a quiescent disk
for sufficiently porous grains, if the effective temperature due to accretion power is a 
significant fraction of the actual temperature several scale heights above the midplane.
Photophoretic dust levitation is driven by the optically thick vertical flux of energy resulting from the dissipation of accretion energy deep in the disk.
This flux can be parameterized in terms of the accretion power's contribution to the effective temperature of the disk as:
\begin{equation}
P(z)= \sigma_{SB} T_\mathrm{e,acc}^4(z).
\end{equation}
where we denote this contribution as $T_\mathrm{e,acc}$ to explicitly note that it results from accretion power, not irradiation.
At any given height $h$ in the disk, $2F(z)$ is the vertically integrated accretion power released below that height $|z|<h$, where the factor
of $2$ accounts for the two sides of the disk.
In this parameterization, $T_\mathrm{e,acc}$ is the effective temperature that the disk would have if the accretion power released below $|z|<h$ were the only
energy source for the disk, and can be compared to the irradiation temperature.

Expressing the vertical  energy flux $E$ in terms of the optically thick flux using Equation~(\ref{eq_Itheta}) as 
\begin{align}
E = \int_0^{\pi}\int_0^{2\pi} I(\theta) \cos \theta \sin \theta  d\phi d\theta = \frac{16}{3} \sigma_{SB} \Gamma T_g^4,
\end{align}
we find the the resulting temperature gradient must be
\begin{align}
\Gamma = \frac{3}{16} \frac{T_{e}^4}{T_g^4}.
\end{align}
We will seek to balance the  photophoresis force on a sphere Equation~(\ref{eq_sphere_force}) 
against the gravitational force on a dust grain of mass $m_d$ given as
\begin{align}
{ F}_g = -m_d z \Omega^2 \ez .
\end{align}
Combining the above three equations with the equations for an MMSN we find that gravity and photophoresis
balance at a critical effective temperature  $T_\mathrm{e,crit}$ when
\begin{align}
 \frac{\pi}{6} \frac{ k_B }{\mu m_H}  \alpha \frac{a^3}{k}   \rho_g \sigma_{SB} T_\mathrm{e,crit}^4 &
 \left( 1  +4 \sigma_{SB} T_g^3  \frac{a}{k}  + \Upsilon T_g^{1/2}  \frac{a}{k} \right)^{-1} \nonumber\\
 & = \frac{4 \pi \phi \rho_0 a^3}{3} z \Omega^2.
\end{align}
Using Equation~(\ref{Omega_Eq}) we can simplify the above to
\begin{align}
 \frac{1}{8}    \frac{\alpha}{k}   \rho_g \sigma_{SB} T_e^4 &
 \left( 1  +4 \sigma_{SB} T_g^3  \frac{a}{k}  + \Upsilon T_g^{1/2}  \frac{a}{k} \right)^{-1}  \nonumber\\
 & = \phi \rho_0 z \frac{T_g}{H^2}.
\end{align}
Rearranging terms and using Equations~(\ref{rho_g_equation}) and (\ref{Sigma_g_equation}) we come to
\begin{align}
\left(\frac{T_\mathrm{e,crit}}{T_g}\right)^4 = \frac{\sqrt{2 \pi} 8}{\sigma_{SB} T_g^3 } &\frac{\rho_0}{\Sigma_g} \frac{k \phi}{\alpha} 
 \left(\frac zH e^{+z^2/2H^2}\right) \nonumber\\
 &\left( 1  +4 \sigma_{SB} T_g^3  \frac{a}{k}  + \Upsilon T_g^{1/2}  \frac{a}{k} \right),
 \label{TecritTratio}
\end{align}
where we note that $\Upsilon$ depends on both $\Sigma_g$ and $z/H$ through its dependence on $\rho_g$.

In Figure~\ref{figlevitation} we plot the critical effective temperature $T_{e,crit}$ such that Equation~(\ref{TecritTratio})
is satisfied in an MMSN background. Our parameters are $R, z/H, \phi$, and $a$ (and hence dust particle density $m_d$). 
We include contours
of $\log(T_\mathrm{e,crit}/T_g)=-0.5$, at which level the accretion power would only be $1\%$ of the background disk luminosity. That
represents a very quiescent disk almost fully heated through irradiation, and is a conservative estimate for a level
of accretion power that must be considered in any protoplanetary disk model.

In the top left panel of  Figure~\ref{figlevitation} 
we show that for $\phi \lesssim 0.3$, porous dust grain up to approximately $1$\,mm in radius are highly subject
to photophoretic levitation up to at least $z/H=1$ at $R=1$\,AU; demonstrating the basic importance of considering
photophoresis in terrestrial planet forming regions. In the top right panel we show that highly porous $\phi \lesssim 0.12$ grains are highly subject
to photophoretic levitation out to at least $R=2$\,AU, and most likely into the asteroid belt region $R=2-3$\,AU; however, photophoresis
requires significant levels of accretion power to play a major role farther out.
In the bottom left panel we show the levitation profile for a dust grain with mass comparable to chondrules found
in the LL3.0 chondrite Semarkona, with a radius $a \sim 0.25\ \mathrm{mm}$ \citep{2014arXiv1408.6581F},
demonstrating that chondrule precursors are likely to have been levitated away from the midplane.
Finally, the bottom right panel shows that highly porous dust grains would have been levitated well above
the midplane, to about $z/H \gtrsim 2$, until they grow past the $100\,\mu$m mark, at which point
photophoretic levitation begins to be less effective (although it can still play a significant role for particles with $a\gtrsim1 \ \mathrm{mm}$).

In aggregate, Figure~\ref{figlevitation} demonstrates that photophoretic levitation is expected to play a major role in vertically
distributing highly porous, collisionally grown dust in the terrestrial planet forming region of protoplanetary disk, out to the current location of
the asteroid belts. However, as dust processing continues, the larger and more compacted grains would decouple from the photophoretic
force; which could only play a role during accretion events such as FU Orionis outbursts, when we would expect
$T_e > T$. In particular, we note that chondrule precursors are likely to have been levitated above the midplane.

\section{Conclusions}\label{sec_conclusions}

We have derived the general theory for photophoresis force on an opaque, homogeneous,
spherical particle with either one or two layers in a dilute, optically thick medium.
This analysis is simple and general, making no assumptions specific for protoplanetary disks.
In itself, the theory presented here applies equally well to any astrophysical 
or terrestrial system with the same basic criteria.

The theory developed here for spherical grains shows that the effects are most pronounced for the lowest conductivity grains.
However, those grains have that property because they are very porous. The assumptions of opaque
spherical particles, used here and in \citet{2012A&A...545A..36L}, 
must break down at some point for sufficiently porous grains. Improving on this state of the art is 
clearly important future work.

In protoplanetary disks, photophoresis driven by accretion processes dissipating energy into heat
can have dynamical impacts for highly porous grains, including levitating chondrule precursors mass
objects above the midplane. This effect would rapidly decay as collisions or thermal processing compactify
the grains.
Photophoresis as a levitation effect in protoplanetary disks may act in concert with other mechanisms, 
such as the drag due to a disk wind emanating from deep in the disk \citep{2015arXiv150503704M}.
{ However, here it important to note that the theory derived in this work applies to opaque grains, 
which also implies that to feel a photophoretic force 
they must have a size significantly greater than the wavelength of the thermal radiation in the disk. }

The analysis of photophoresis driven dust motion in Section~\ref{sec_applications} treated only silicate dust.
However grains made of other, such as graphite/organics, ices, or composites, all with different thermal properties, also populate
 protoplanetary disks; and can be expected to follow different protophoresis driven motions.
This work motivates basic experiments to determine the appropriate parameters for these more diverse grain compositions.

\acknowledgments
We acknowledge enlightening conversations with Christophe Loesche and Gerhard Wurm on 
photophoresis driven by temperature fluctuations which lead to 
the concept behind this work. 
We thank the anonymous referee and Farrukh Naumann, and Melissa McClure for useful 
comments which improved the manuscript.
The research leading to these results has received funding from the 
People Programme (Marie Curie Actions) of the 
European Union's Seventh Framework Programme 
(FP7/2007-2013) under REA grant agreement 327995 and ERC grant agreement 306614 (CPM), and 
NASA OSS grant NNX14AJ56G  (AH).

\appendix

\section{Numerical solution to single-layer sphere heat transfer}
\label{app_numericalsolution} 
To solve Equation~(\ref{eq_solid_bc}) numerically we define the residual $R$ 
as a function of the vector ${ A}$ of coefficents in the Legendre polynomial approximation:
\begin{align}
R({ A}) &\equiv k\left. \frac{\partial T}{\partial r}\right|_{a}  +\sigma_{SB}T^4  -I_c(\theta)  - I_n(\theta).
\end{align}
We then solve for ${ A}$ which minimizes the residual using a Levenberg-Marquardt least-squares algorithm. 
Typically we use 32 Legendre polynomials.
To check the accuracy of the linear solution which has been expanded in $\eta\ll 1$ 
we wish to isolate the primary nonlinear term,  with $T^4$, 
so we set the conductive cooling flux $I_c=0$.
Then, to analyze the relative error induced by the linear approximation,
\begin{align}
\left(\frac{A_1 a}{A_0}\right)_\text{linear} \big/ \left(\frac{A_1 a}{A_0}\right)_\text{numerical} -1,
\end{align}
we need to only vary the parameters $\sigma_{SB} a/k$ and $T_g$. 
For dust motion in a protoplanetary disk, 
$T_g \in (0\,\text{K},10^4\,\text{K}]$, $\sigma_{SB} a/k \in [10^2,10^9]$, and $\Gamma \in [0,10^{-3}]$.
Over these ranges we find that the relative error is
always less than $10^{-6}$, so the linear approximation to $A_1$ is excellent.

\section{Linear solution to double-layer sphere heat transfer}
\label{app_lineardoublesolution} 

In the core and mantle layers, the temperature can be given in the series expansions:
\begin{align}
T_c(\theta) &= \sum_n \left( C_n r^n  \right) P_n(\cos(\theta)), \label{legendre_Tc}\\
T_m(\theta) &= \sum_n \left( A_n r^n +  B_n r^{-n-1} \right) P_n(\cos(\theta)). \label{legendre_Tm}
\end{align}
The boundary conditions at the core radius $r_c$ are
\begin{align}
k_c \left. \frac{\partial T_c}{\partial r} \right|_{r_c} &= k_m \left. \frac{\partial T_m}{\partial r}\right|_{r_c},\\
T_c(r_c,\theta) &= T_m(r_c,\theta),
\end{align}
where the second condition assumes that the interface has perfect conduction.
At the outer, mantle plus core radius $a$, the boundary condition is
\begin{align}
k_m \frac{\partial T_m }{\partial r}  +4\sigma_{SB}T_m^4 + \Upsilon T_g^{1/2} (T_g-T_2)  = I_n(\theta).
\label{eq_rm_bc}
\end{align}
Expanding the second left hand side term of Equation~(\ref{eq_rm_bc}) and substituting the expansions in
Equations~(\ref{legendre_Tc}) and (\ref{legendre_Tm}) into the 
$r_c$ and $a$ boundary conditions leads to the following relations for the $n=1$ coefficients: 
\begin{align}
&k_c C_1 = k_m(A_1 - 2 r_c^{-3} B_1), \label{eq_dblbcsys1}\\
&r_c C_1 = r_c A_1 + r_c^{-2} B_1,\\
&k_m (A_1 - 2a^{-3} B_1) + 4\sigma_{SB} T_0 ^3(a A_1 + a^{-2} B_1) 
\nonumber \\
&- \Upsilon T_g^{1/2} (-a A_1 - a^{-2} B_1)  = \sigma_{SB} T_g^4 \left(-\frac{8}{3} \Gamma\right).\label{eq_dblbcsys3}
\end{align}
The 
$n=0$ terms yield $T_0=T_g$ and $C_0  = A_0 +B_0r_c^{-1}$.
Solving the system of equations~(\ref{eq_dblbcsys1})--(\ref{eq_dblbcsys3}) yields the 
$n=1$ coefficients in the $T_m$ series:
\begin{align}
A_1 a + B_1 a^{-2}  =& -\frac{8}{3} \sigma_{SB} T_g^4 \Gamma  \frac{a}{ k_m}
  \left[(1-b^3)+\frac{k_m}{k_c} (2+b^3)  \right] \nonumber\\
&  \times 
\bigg[
    2  \frac{k_m}{k_c} (1-b^3) + \left( 1 +2 b^3  \right)
     \nonumber\\
 & \quad + \frac{a}{k_m} \left(  (1 - b^3)   
        +  \frac{k_m}{k_c} (2+b^3) \right)
         \nonumber\\
 & \quad \times \left( \Upsilon T_g^{1/2}  +4\sigma_{SB}T_g^3\right)
    \bigg] ^{-1}, \label{n1coefficients}
\end{align}
where $ab=r_c$. 
Using Equations~(\ref{legendre_Tm}) and (\ref{n1coefficients}), the approximate photophoresis force integral (Equation~\ref{eq_photforce_approx}) yields
\begin{align}
{ F}_p &\approx -\frac{\pi}{3} \alpha \frac{p}{T_g} a^2 (A_1 a +B_1 a^{-2}) \ez.
\label{eq_photoforce_a1_b1}
\end{align}
This completes the solution for the photophoresis force on the double-layered sphere.

\end{document}